\documentclass[aps,prl,twocolumn,superscriptaddress,showpacs]{revtex4}

\usepackage{amsmath,float,graphicx,epsfig,subfigure}

\begin{document}

\title{Quantum vortices in optical lattices}
\author{P. Vignolo}
\affiliation{NEST-CNR-INFM and Scuola Normale Superiore, I-56126 Pisa, Italy}
\affiliation{INFN, largo B. Pontecorvo 3, I-56127 Pisa, Italy}
\author{R. Fazio}
\affiliation{International School for Advanced Studies (SISSA),
                via  Beirut 2-4,  I-34014 Trieste, Italy}
\affiliation{NEST-CNR-INFM and Scuola Normale Superiore, I-56126 Pisa, Italy}
\author{M.P. Tosi}
\affiliation{NEST-CNR-INFM and Scuola Normale Superiore, I-56126 Pisa, Italy}

\begin{abstract}
A vortex in a superfluid gas inside an optical lattice can behave as a 
massive particle moving in a periodic potential and exhibiting quantum 
properties. In this Letter we discuss these properties and show that the 
excitation of vortex motions in a two-dimensional lattice can lead to 
striking measurable changes in its dynamic response. It would be possible 
by means of Bragg spectroscopy to carry out the first direct measurement 
of the effective vortex mass, the pinning to the underlying lattice, and 
the dissipative damping.
\end{abstract}
\pacs{05.30.Jp; 74.78.w; 74.81.Fa}
\maketitle
The understanding of the static and dynamical behavior of vortices has 
been crucial to describe numerous different situations in superfluids 
ranging from liquid Helium to high-temperature 
superconductors~\cite{volovik,tinkham}. These defects can be created 
by means of an applied magnetic field in superconductors and by putting 
the sample into rotation in superfluid Helium, or they can be thermally 
excited in low-dimensional systems where the unbinding of vortex-antivortex 
pairs is at the core of the Berezinskii-Kosterlitz-Thouless transition. 
Low-dimensional superconductors and in particular Josephson Junction 
Arrays (JJAs) have been for many years a natural playground for studying 
classical and quantum properties of vortices~\cite{fazio}. A vortex 
in a JJA behaves as a massive particle moving in a periodic potential and subject 
to dissipation~\cite{eckern}, and under appropriate conditions vortices 
can show quantum properties such as interference or tunneling. 
Among the most interesting experiments performed with vortices in 
JJAs we mention the observation of ballistic motion~\cite{zant}, 
the measurement of the Aharonov-Casher effect for a vortex 
going around a charge~\cite{elion}, and the Mott-Anderson insulator 
of vortices~\cite{oudenaarden}.

Optical lattices for atomic gases, which currently are under intense 
investigation~\cite{minguzzi03,lewenstein,Morsch}, can behave as 
tunneling junction arrays. In this Letter we analyze vortex excitations 
in an optical lattice and show that a superfluid gas in an optical lattice 
offers a unique opportunity for a 
{\it direct measurement of vortex properties} 
(such as the mass, the coupling to its environment or the pinning potential) 
{\it via} a Bragg spectroscopy experiment. 
This is in contrast to the JJA case, 
where only indirect measurements based on transport properties are available. 
The Bragg spectroscopy technique~\cite{Ketterle1,Ketterle2,gardiner}  
has been appealed to for a variety of experiments on ultra-cold atomic gases, 
and in optical lattices has been considered for a measurement of the 
excitation spectrum of a Bose gas in the Mott-insulator phase~\cite{rey},
and of its coexistence with a superfluid phase in a dishomogeneous 
cloud~\cite{batrouni}.

We consider a Bose gas in the superfluid phase inside a 
lattice~\cite{jaksch,greiner}, 
in a regime where the hopping and the on-site repulsion between the bosons 
are competitive. Quantum fluctuations due to the interplay of the local 
repulsions and of the hopping have dramatic consequences for vortex dynamics. 
In this case a vortex behaves as a macroscopic quantum particle, moving in a 
periodic potential with a mass that we evaluate and show to be directly 
measurable by Bragg spectroscopy. At variance from other recent studies of 
vortices in frustrated optical 
lattices~\cite{zoller, mueller,demler,polini,palmer,osterloh},
we discuss the {\it dynamical} properties of an {\it individual} vortex. 
In order to achieve this regime one can either apply a very low 
frustration by means of a rotation of the lattice~\cite{tung} 
generating only a few and very weakly interacting vortices, or 
create a vortex excitation by means of phase 
inprinting~\cite{matthews99,claus}.
Of particular relevance is the very recent observation of vortex pinning 
in co-rotating optical lattices by Tung {\it et al.}~\cite{tung},
which indicates that what we propose here is within reach of experimental 
verification.

\paragraph{The model - }
We consider a Bose gas at zero temperature inside a square lattice with 
lattice constant $a$ and $N_s$ lattice sites. We assume that the system can 
be described by a single-band Bose-Hubbard Hamiltonian~\cite{fisher}
\begin{equation}
        H=-\frac{J}{2}\sum_{\langle ij\rangle} \;
        \hat b^{\dagger}_{i}\hat b_{j}+ \mbox{H.c.}+
 U\sum_{i}\hat n_{i}(\hat n_{i}-1) - \mu \sum_{i} \hat n_{i},
        \label{bh}
\end{equation}
where $\hat b_i^{\dagger}$ and $\hat b_i$ are the creation and annihilation 
operators for a boson on the $i$-th site and $\hat n_i =\hat b^{\dagger}_i
\hat b_i $ is the
number operator. The coupling constant $U$ describes the local interaction
between bosons, $\mu$ is the chemical potential, and $J$ the
matrix element for hopping between nearest-neighbors sites. 
The on-site interaction energy and the hopping energy  are given by 
$U=g\int d{\bf r}\,|w_0({\bf r}-{\bf R}_i)|^4$,
$J=-(\hbar^2/2m)^{-1}\int d{\bf r}\,w_0^*({\bf r}-{\bf R}_i)\nabla^2
w_0({\bf r}-{\bf R}_j)
$, in terms of the Wannier function $w_0({\bf r})$ (${\bf R}_i$ is the coordinate of the 
 $i$-th site).
Here 
$g=4\pi\hbar^2a_{sc}/(\sqrt{2\pi}m l_\perp)$ is the repulsive interaction 
strength in the case where the transverse size $l_\perp$ of the  
lattice is larger than the scattering length $a_{sc}$.

If the 
average number of bosons per site is much larger than one, 
the field operators can be approximated as
$\hat b_i\simeq\sqrt{\bar n}\exp{(i\hat\phi_i)}$, with $\hat\phi_i$ being
the phase operator on the $i$-th site.
The Bose-Hubbard model can be recast into the quantum phase Hamiltonian 
\begin{equation}
\hat H=-J\bar n\sum_{\langle i,j\rangle}\cos(\hat \phi_i-\hat \phi_j)+
U\sum_i \delta\hat n_i^2 - \tilde{\mu}\sum_i \delta\hat n_i
 \label{qpm}
\end{equation} 
where $\tilde{\mu} = 2U -\mu-1$.
The number operator has been expressed in terms of the fluctuations around its 
average value $\bar n$, $\hat n_i=\bar n+\delta\hat n_i$. 
The number fluctuation operator and the phase are canonically conjugate 
variables,
$[\delta \hat n_i, e^{\pm i\phi_j}]=\delta_{ij}e^{\pm i\phi_j}$.
The regime that we consider throughout this work is $J \bar n\gg U$: 
the system is deep in the superfluid region, but quantum fluctuations 
are present and play a crucial role in the vortex dynamics. 

\paragraph{Vortex properties - }
The presence of a static vortex inside the lattice 
can be described to a good approximation by a phase 
distribution of the boson field given by
\begin{equation}
\phi_i= {\rm arctan}\left(\dfrac{y_i-y}{x_i-x}\right)\,,
\label{phases}
\end{equation}
where $x,y$ are the coordinates of the center of the vortex.  
Deep in the superfluid regime and at temperatures much lower than 
$J \bar{n}/K_B$,
phase rigidity ensures that again to a good approximation,
a moving vortex can 
still be described by Eq.~(\ref{phases}) but with a time-dependent  
position of the vortex 
center. 
The existence of a vortex mass can be understood 
qualitatively by noting that if a vortex moves of a distance of the order 
of $a$ in a time $\delta t = a/v$, $v$ being its velocity, 
the phase difference $\delta\phi_{ij}$ at each  bond changes in time as
$\delta\phi_{ij} =  \phi_{ij}(t+a/v)-  \phi_{ij}(t)$. 
Due to the commutation relation between the number and 
phase operators, a time-dependent phase leads to a contribution to the energy
which is quadratic in the vortex velocity (see the second term of the r.h.s of 
Eq.(\ref{qpm})). The problem of calculating the vortex mass can then be reduced to 
find the phase differences across junctions at times
$t$ and $t+a/v$. 

An effective action for a vortex in a lattice can then be obtained by 
inserting Eq.~(\ref{phases}) in the Bose-Hubbard model in Eq. (\ref{qpm}) 
and then expressing the resulting 
action in terms of the vortex coordinates 
${\bf r}(t)=\{x(t),y(t)\}$~\cite{eckern}. 
The on-site repulsion term provides a kinetic energy term 
$T=(M_v/2)\dot{{\bf r}}^2$, where the vortex mass in a lattice of size $L$
is
\begin{equation}
M_v=\frac{\sqrt{2\pi}l_\perp w^2}{4 a_{sc} a^2}m\,\ln(L/a).
\label{mvortice}
\end{equation}
The vortex mass thus scales linearly with the boson mass, increases with  
the width $w$ of the Wannier function, and decreases with
the scattering length.

\begin{figure}[H]
\begin{center}
\epsfig{file=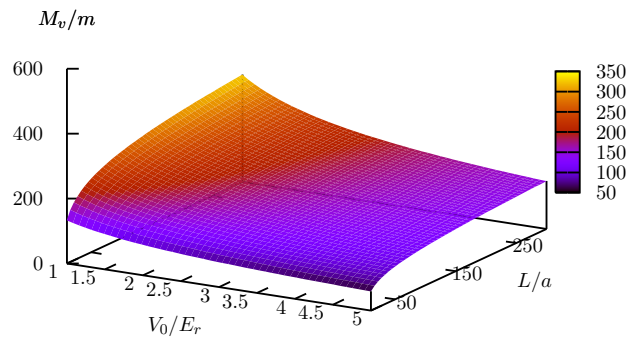,width=0.95\linewidth}
\caption{The vortex mass $M_v$ (in units of the boson mass $m$) as a function
of $V_0/E_r$ and $L/a$, for the case $a=426$ nm, 
$l_\perp=5\,\mu$m, and $a_{sc}=5.5$ nm.}
\label{fig_vortex}
\end{center}
\end{figure}

Let us consider for an
illustration the $^{87}$Rb lattice realized by
Greiner and coworkers~\cite{greiner}, in the case $V_0=4E_r$ where
$V_0$ and $E_r$ are the well depth of the optical lattice and the recoil 
energy respectively.
For such a system in 2D we have $a=426$ nm, $w\simeq 96$ nm, 
$l_\perp=5\,\mu$m, $a_{sc}=5.5$ nm and $L=75$ $\mu$m, and the
vortex mass is
\begin{equation}
M_v \simeq 29 m\ln(L/a)\simeq 150 m\simeq 2.2\times 10^{-20}{\rm gr}.
\end{equation}
The behavior of the vortex mass as a function of the potential well 
depth (which affects $w$)
and of the size of the lattice is depicted in Fig. \ref{fig_vortex}.

The effective potential seen by the vortex has been numerically evaluated 
in the context of JJAs and, in the case of a 
vortex moving in the $x$ direction inside a
large two-dimensional array 
the effective potential is periodic~\cite{lobb},
\begin{equation}
U_v(x)=0.1J\bar{n}[\cos(2\pi x/a)-1] \;.
\label{period}
\end{equation}
In the presence of a vortex the whole array thus behaves as 
a macroscopic particle of mass $M_v$ moving in a periodic potential. 
For such a macroscopic object one has to also take into 
account the interaction with the environment, and the main
source of damping is due to the interaction with the long-wavelength 
phase modes that are excited during vortex motion. This 
damping has been analyzed in detail in the context of JJAs
(see for example~\cite{uli,anne}). 
In this Letter for simplicity we shall just comment on
how our results are modified in accounting for dissipation.

\paragraph{Dynamic structure factor  - }
We turn to a calculation of the dynamical response of the Bose gas
inside a lattice, both in the absence and in the presence of a vortex.
The central quantity of interest is the dynamic structure factor,
which in a tight-binding scheme takes the form
\begin{equation}
S({\bf q},\omega)= \int dt e^{i\omega t}\sum_{i,j} 
e^{-i{\bf q}\cdot({\bf R}_i-{\bf R}_j)}
\langle \delta \hat n_i(t)\delta \hat n_j(0)\rangle.
\label{Somega}
\end{equation}
This spectrum can be measured in experiments
of Bragg spectroscopy~\cite{Ketterle1,Ketterle2}:
two probe laser beams, with frequencies $\omega_1$ and 
$\omega_2=\omega_1+\omega$ and wave-vectors ${\bf k}_1$ and 
${\bf k}_2={\bf k}_1+{\bf q}$,
scatter on the boson gas and the spectrum
measures the probability of momentum transfer $\hbar{\bf q}$ at energy 
$\hbar\omega$~\cite{gardiner}.
The {\it f}-sum rule gives the first spectral moment $M_1({\bf q})$
as $M_1\equiv\int S({\bf q},\omega)\omega\,d\omega=
\frac{1}{2\hbar}
\langle 0|[\delta \hat n_{\bf q},
[\hat H,\delta \hat n_{\bf q}^\dagger]|0\rangle\,,$
$\delta \hat n_{\bf q}$ being the Fourier transform of $\delta \hat n_i$.

We first consider the case in which no vortex is present.
Inside the superfluid regime ($J\bar n \gg U$), it is enough to 
consider long-wavelength phase fluctuations, as described by 
expansion of the cosine in the phase Hamiltonian up to second order. In this 
limit the Hamiltonian is easily diagonalized in Fourier space 
by means of the transformations
$\hat\phi_{\bf k} = [U N_s/(\hbar\Omega_{\bf k})]^{1/2}\left(
\hat a_{\bf k}+\hat a_{-{\bf k}}^\dagger\right)/\sqrt{2}$ and 
$
\hat n_{\bf k} = (N_s\hbar\Omega_{\bf k}/U)^{1/2}\left(
\hat a_{\bf k}-\hat a_{-{\bf k}}^\dagger\right)/i\sqrt{2}
$,
with the result
\begin{equation}
\hat H = \sum_{{\bf k}
\in{\rm BZ}}\hbar \Omega_{\bf k}
\left(\hat a_{\bf k}^\dagger\hat a_{\bf k}+\frac{1}{2}\right)
\label{harm}
\end{equation}
where
$\Omega_{\bf k}^2=(2J\bar nU/\hbar^2)[2-\cos(k_xa)-\cos(k_ya)]
$. Here the quasi-momentum ${\bf k}=(k_x,k_y)$ is inside the first 
Brillouin zone.
By taking into account the time dependence of the particle number 
fluctuation operator dictated by Eq. (\ref{harm}), it is straightforward 
to obtain the dynamic structure factor as
\begin{equation}
S({\bf q},\omega)= 
\frac{\hbar\Omega_{\bf q}}{2 U} \delta (\omega-\Omega_{\bf q})\,.
\label{eurekabis}
\end{equation}
The physical interpretation of Eq. (\ref{eurekabis}) is clear: 
small-$q$ absorption occurs at a frequency corresponding to the 
dispersion relation of the Goldstone sound mode in the lattice.
In this case
$M_1({\bf q})=(J\bar n/\hbar) [2-\cos(q_x a)-\cos(q_y a)]$.

The presence of a vortex can induce, besides changes in the sound wave 
spectrum, specific contributions associated with excitations
of vortex motions.
Several different situations can be envisaged for the latter,
but here we discuss the interesting
case in which the hopping parameter $J$ is sufficiently large that the 
vortex is pinned to a minimum of the periodic potential given by 
Eq. (\ref{period}). 
In this case the vortex dynamics is associated with small oscillations around 
its equilibrium position and can be described by the harmonic oscillator 
Hamiltonian
\begin{equation}
H_v=\frac{1}{2}M_v\dot{\bf r}^2+\frac{1}{2}M_v\Omega_v^2{\bf r}^2,
\label{lowe}
\end{equation}
where we have defined $\Omega_v=(0.1J\bar{n}/M_v)^{1/2}2\pi/a$. 
By noting that
\begin{equation}
\langle \delta \hat n_i(t_1)\delta \hat n_j(t_2)\rangle=
\frac{U^2}{\hbar^2}
\langle \delta \hat{\dot\phi}_i(t_1)\delta \hat{\dot\phi}_j(t_2)\rangle \;,
\end{equation}
using the expression given in Eq. (\ref{phases}) for the phase distribution 
 and performing an average over the vortex degrees 
of freedom with the help of Eq. (\ref{lowe}), it is possible to 
write the contribution of the vortex to the dynamic structure factor as
\begin{equation}
S_v({\bf q},\omega)=
\frac{\hbar^2\Omega_v}{U^2}
\frac{4\pi^2 \hbar}{M_v a^4 q^2}\,
\delta(\omega-\Omega_v)\,.
\label{swansea}
\end{equation}
Instead of a ${\bf q}$-dependent resonance as in Eq.(\ref{eurekabis}), the 
Bragg scattering acquires a resonance at a well defined frequency $\Omega_v$ indicating 
that the whole lattice responds collectively having the properties of a single 
macroscopic particle, the vortex. This is the main result of this paper.  
Under the conditions specified above, the
presence of a vortex induces a resonance at a frequency
that allows access to the vortex mass. Let us remark that this
resonant behavior is related to the presence of the lattice {\em and} to 
the existence of quantum fluctuations originating from the local repulsion.
The Bragg spectrum
of a vortex in a Bose-Einstein condensate is otherwise determinated by
a dispersion relation~\cite{gardiner}. 
The peculiar dependence of the spectral strength in Eq. (\ref{swansea}) 
on the transferred momentum  $q$ is 
due to the coupling between the exciting radiation and the lattice: 
at low momentum all phases in  the lattice are excited and the dynamic response is enhanced. 
We finally should comment on the fact that Eq.~(\ref{swansea}) does not 
fulfill the $f$-sum 
rule: this should come as no surprise, as this expression is valid only 
at low energy.

The coupling to long-wavelength phase 
fluctuations provides the main dissipation mechanism for the 
vortex motions~\cite{uli,anne,sonin}. To a first approximation this results 
in Ohmic damping on the vortex. 
In the presence of dissipation the delta function in the dynamical response 
is smeared and acquires a finite width proportional to the 
dissipation strength.

In summary, in this Letter we have discussed some main aspects of
quantum dynamics of a vortex in an atomic superfluid gas inside an
optical lattice. 
We have specifically considered the situation in which the vortex is 
pinned by the lattice potential and only executes small oscillations
around its energy minimum. In this case
Bragg spectroscopy should allow a measurement of the vortex mass.
One can envisage other situations in which to study vortex dynamics. 
Experiments on quantum tunneling/coherence  of vortices seem to be  
within experimental reach. 

We acknowledge fruitful discussion with C. Bruder and M. Polini. This work 
has been supported by MIUR-PRIN and by EC through grants EUROSQIP and RTNNANO.

\end{document}